\documentclass[prl,twocolumn,showpacs,preprintnumbers,amsmath,amssymb]{revtex4}

\usepackage{graphicx}

\usepackage{dcolumn}

\usepackage{bm}

\usepackage{subfigure}

\begin{document}

\title{First-principles calculations of spin and angle-resolved resonant 
    photoemission spectra of Cr(110) surfaces at the 2$p$ - 3$d$ resonance} 

\author{F. Da Pieve$^{1}$} 

\author{P. Kr\"uger$^{2}$}

\affiliation{
$^1$ ALGC, Vrije Universiteit Brussel, Pleinlaan 2, 1050 Brussels, Belgium.\\
$^2$ ICB, UMR 6303 CNRS - Universit\'e de Bourgogne, F-21078 Dijon, France.}

\date{\today}


\begin{abstract}
A first principles approach for spin and angle resolved resonant photoemission is
developed within multiple scattering theory and applied to a Cr(110) surface at the
2$p$-3$d$ resonance.  The resonant photocurrent from this non
ferromagnetic system is found to be strongly spin polarized by
circularly polarized light, in agreement with experiments on antiferromagnetic and magnetically disordered systems.
By comparing the antiferromagnetic and Pauli-paramagnetic
phases of Cr, we explicitly show that the spin polarization of the
photocurrent is independent of the existence of local magnetic moments, solving a long-standing debate on the origin of such
polarization. New spin polarization effects are predicted for the paramagnetic phase even with unpolarized light, opening new directions for full mapping of spin interactions in macroscopically non magnetic or nanostructured systems.
\end{abstract}









\maketitle



In recent years, the theoretical description of absorption/photoemission
spectroscopy in the X-ray region has been boosted
by the merge of density functional theory (DFT) with many body approaches 
such as dynamical mean field theory
\cite{sipr,braun}, many body perturbation theory
\cite{olovsson,vinson,blaha} and by the development of time-dependent DFT
\cite{joly}. However, second order processes, like resonant
inelastic X-ray scattering (RIXS) and resonant photoemission (RPES), remain a major challenge for theory. For RPES, existing approaches are semiempirical \cite{mishra,tanakamult,degroot,kotani}, based on a well defined two-holes final state and on small clusters, and thus do not take into account the delocalization of intermediate states, the 
bandstructure of the system and multiple scattering effects in the propagation of photoelectrons.

The huge experimental output from RPES on correlated materials \cite{mishra,morscher,rich,dilmagnoxides,tjeng,jes117,sinkovic} and the intriguing quest for a determination of local magnetic properties put forward by pioneering experiments ~\cite{tjeng,jes117,sinkovic} call for advancements in the theoretical description of this spectroscopy.  In experiments on CuO and Ni, it was shown that the RPES photocurrent with circular polarized 
light is spin polarized in antiferromagnets~\cite{tjeng,jes117} 
and Curie paramagnets~\cite{sinkovic}.
It was claimed that a specific combination of spin resolved spectra provides a direct measure of the local 
magnetic moments~\cite{tjeng,jes117,sinkovic}. The issue is of fundamental importance in the search for a tool to access the local magnetic properties in antiferromagnetic, magnetically disordered and/or nanostructured systems at their
crossover with the transition temperature.
The interpretation was however rejected on the basis of symmetry analysis \cite{vdlcomment}, but explicit calculations predicting the lineshape and intensity of such fundamental signal are still lacking and remain highly desirable.

In this letter, we present the first ab-initio method for RPES
in solids, based on a combined formulation within the real space multiple scattering 
(RSMS) approach~\cite{rehr00,sebilleau06} and DFT, and its application to Cr(110) at the 2$p$-3$d$ resonance.
By comparing the antiferromagnetic (AFM) and Pauli-paramagnetic (PM) 
phase of Cr, we solve the long-standing debate about the possibility to 
determine local magnetic moments in macroscopically non magnetic systems 
by means of spin resolved RPES with circular polarized light. 
New interesting effects in the PM phase by unpolarized light suggest that other mechanisms are active and could be exploited for mapping the origin of the different spin polarization (SP) components in paramagnets and magnetically disordered systems.

{\it Theoretical formulation.}
The cross section for valence band photoemission to a final state $|\underline{v},k\rangle$, where $\underline{v}$ denotes a valence band hole 
and $k$ a photoelectron state, is given by
\[
I(\omega,q,k) = 
\sum_v |T_{kv}(\omega,q)|^2 \delta(\epsilon_k - \epsilon_v - \hbar\omega)
\]
where $\hbar\omega$
and ${q}$ are the photon energy and polarization. Here the independent particle approximation has been assumed (i.e., all many-electron eigenstates are single Slater
determinants corresponding to the same effective one-electron
hamiltonian). According to the Heisenberg-Kramers formula~\cite{tanaka},
the transition matrix element $T_{kv}(\omega,q)$ is the sum of a direct 
and a resonant term.
In the latter, photon absorption leads to an
intermediate state $|\underline{c},u\rangle$, with a core 
hole $(\underline{c})$ and an electron in a formerly unoccupied state $|u\rangle$, which decays to the final state $|\underline{v},k\rangle$
through a participator Auger process~\cite{tanaka,janowitz92}. To lowest order in the autoionization process, the transition
matrix element is given by
\begin{equation}\label{eqT}
T_{kv}(\omega,q) = \langle k|D_q|v\rangle 
+ \sum_{cu} \frac{ \langle kc|V (| vu\rangle-|uv\rangle)}{
\hbar\omega + \epsilon_c-\epsilon_u-i\Gamma} \langle u |D_q| c\rangle
\end{equation}
where $D_q$ is the dipole operator, $V$ the Coulomb operator and $\Gamma$ the width of 
the intermediate state. Spectator Auger decay leads to different, namely two-hole
final states and is not considered here. Participator and spectator channels can in principle be
separated experimentally by using a photon bandwidth smaller than the core-hole lifetime, as they show different photon energy dependence (linear for the participator, and no photon energy dependence for the spectator). Here we focus on the physical effects at the origin of spin polarization and dichroism as well as their directional-dependence in the ``pure'' participator channel.

The RPES intensity can be written in a compact form as
\[
I(\omega,q,k) = \sum_{ijLL'\sigma} 
M^{\omega,q}_{iL\sigma}(k)I^{ij}_{LL'}(\epsilon_v,\sigma)
M^{\omega,q}_{jL'\sigma}(k)^*
\]
Here, $i,j$ label atomic sites, $L\equiv (lm)$ angular momentum 
and $\sigma$ spin quantum numbers.
$\epsilon_v=\epsilon_k-\hbar\omega$ is the energy of the valence
hole. The quantity 
$I^{ij}_{LL'}\equiv -\frac{1}{2i\pi}(\tau-\tau^\dagger)^{ij}_{LL'}$  
is the essentially imaginary part of the scattering path operator. It comes from the simplification of the sum over delocalized valence states through the so called optical theorem in RSMS ~\cite{krueger11} and it contains the bandstructure information.
The matrix elements $M^{\omega,q}_{iL\sigma}(k)$
are given by
\[
M^{\omega,q}_{iL\sigma}(k)
= \sum_{jL'} B^*_{jL'}(k) A_{jL',iL}(\epsilon_k\sigma_k, \epsilon_v\sigma)
\]
The $B_{jL'}(k)$ are the key quantities in the RSMS approach and represent the multiple scattering amplitudes of 
the continuum state $k\equiv ({\bf k}\sigma_k)$~\cite{krueger11}.
The matrix elements $A_{jL',iL}(\epsilon_k\sigma_k, \epsilon_v\sigma)$ 
are given by the sum of the direct radiative process ($A^D$), the resonant process 
with direct Coulomb decay ($A^C$) and the resonant process
with the exchange decay ($A^X$), see Eq.~(\ref{eqT}). 
$A^D$ and $A^C$ are site- and spin-diagonal 
($\sim\delta_{ij}\delta_{\sigma_k\sigma}$).
We have 
\begin{eqnarray}
&& A^D =
\langle i\epsilon_k L'\sigma| D_q |i\epsilon_v L\sigma \rangle
\nonumber
\\ && A^C =
-\sum_{j'cL_uL'_u\sigma_u} \int_{E_F}d\epsilon_u 
\frac{I_{L_uL'_u}^{j'j'}(\epsilon_u\sigma_u)}
{\hbar\omega+\epsilon_c-\epsilon_u-i\Gamma} \;\times
\nonumber
\\ &&
\langle i\epsilon_kL'\sigma,j'c|V
|i\epsilon_vL\sigma,j'\epsilon_uL_u\sigma_u\rangle
\langle j'\epsilon_uL_u'\sigma_u|D|j'c\rangle
\nonumber
\\ && A^X =
\sum_{cL_uL'_u} \int_{E_F}d\epsilon_u 
\frac{I_{L_uL'_u}^{ji}(\epsilon_u\sigma_k)}
{\hbar\omega+\epsilon_c-\epsilon_u-i\Gamma} \;\times
\nonumber
\\ &&
\langle j\epsilon_kL_k\sigma_k,ic|V
|j\epsilon_uL_u\sigma_k,i\epsilon_vL\sigma\rangle
\langle i\epsilon_uL_u'\sigma_k|D|ic\rangle
\nonumber
\end{eqnarray}
The sums over unoccupied states $u$ have been again simplified
through the optical theorem.
The exchange term $A^X$ is not strictly site-diagonal
because of the non-locality of the exchange interaction 
together with the delocalized nature of the states~$u$.
In the RSMS approach the Coulomb matrix elements 
$\langle kc|V|vu\rangle$ and $\langle kc|V|uv\rangle$ 
can be exactly developed in one- and two-center terms.
In metallic Cr, the Coulomb interaction is strongly screened. As a result,
two-center terms are by at least one order of magnitude smaller 
than the one-center terms~\cite{note} and have been neglected here.
In general, the 2$p$-3$d$ excited intermediate states might display excitonic 
effects, which could be taken account for with a Bethe-Salpeter description~\cite{olovsson,blaha}. For Cr metal, these effects are quite small because of the large 3$d$ band 
width ($\sim 7$ eV) and efficient metallic screening of the core hole by 
nearly free 4$sp$ electrons, and thus neglected here.

Photoemission spectra from Cr(110) are calculated in RSMS 
with a cluster of 151 atoms (see Fig.~1a) and
self-consistent spin polarized potentials, 
obtained by a scalar relativistic LMTO \cite{lmto}
calculation for bulk Cr in the local spin density approximation.
Except for the 2p core level, all states entering the RPES
 calculation are developed in RSMS. The 2p orbital is obtained
 by solving the scalar relativistic Schr\"odinger equation with
 self-consistent spin-polarized LMTO potentials.
 The 2p$_{3/2}$ spin-orbit coupled states are then constructed using
 standard angular momentum algebra and the spin-orbit coupling
 constant is taken from an atomic calculation~\cite{cowan}. We consider the AFM order of CsCl-type which is a good
approximation to the true spin density wave (SDW) ground state of Cr.
The calculated magnetic moment is 0.74~$\mu_B$ in reasonable agreement with
experiment (0.62~$\mu_B$). At the (110) surface, the transverse SDW
propagates along [100] or [010]~\cite{kfbraun}. Therefore, we take
${\bf e}_z = [001]$ as magnetization and spin-quantization axis throughout this paper.
We also consider the Pauli PM 
state, corresponding to a non-magnetic calculation.
Spin orbit (SO) coupling of the valence and continuum states
is neglected (it is as small as 0.03 eV for Cr-3$d$~\cite{vdlaan91}).

\begin{figure}[!htb]
\parbox{0.5\columnwidth}{
\begin{center}
\hspace*{0.5em} \includegraphics[clip=,width=0.44\columnwidth]{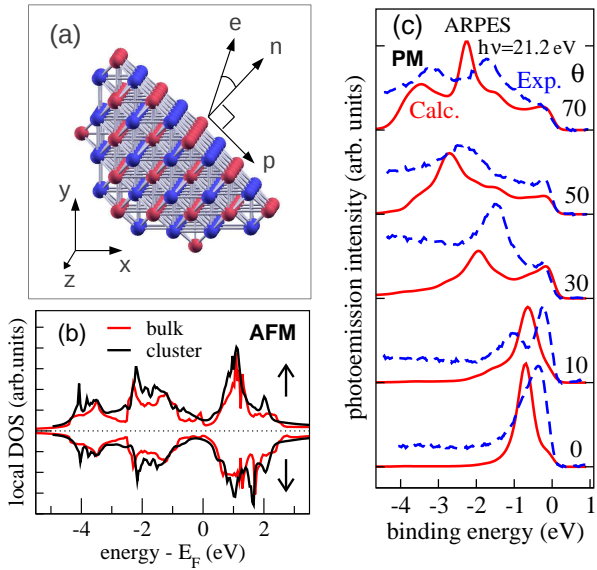}\\
\includegraphics[clip=,width=0.47\columnwidth]{fig1b.eps}
\end{center}
}
\parbox{0.43\columnwidth}{
\begin{center}
\includegraphics[clip=,width=0.39\columnwidth]{fig1c.eps}
\end{center}
}

\caption{
(a) Cr(110) cluster used in the RSMS calculations. 
The two magnetic sublattices of the AFM state are in red and blue.
(b) DOS in the AFM phase for a bulk atom 
(LMTO) and a central atom in the cluster (RSMS).
(c) ARPES spectra from Cr(110) along the $\langle 001\rangle$ azimuth 
for different polar angles $\theta$ with respect to the surface normal. Unpolarized light along the [001] axis was considered. Experimental data from \protect\cite{prb34}.}
\label{fig:conf1}
\end{figure}

\textit{Results.}
The electronic structure of Cr(110) is well accounted for
in the RSMS approach as can be seen from 
the comparison between the local
density of states (DOS) of a Cr atom in the cluster
and of bulk Cr (Fig.~\ref{fig:conf1}b).
Non-resonant angle-resolved photoemission spectra (ARPES)
are shown in Fig.~\ref{fig:conf1}c. Differences with respect to experiments~\cite{prb34} are expected as our approach does not contain local many-body interactions and layer-dependent potentials, which could play a role for a quantitative description of the peak renormalization and dispersion behaviour of the energetic structures \cite{barriga}. However, the main features of the experimental spectra are reproduced in the calculation, confirming that
RSMS provides a reasonably good description of valence band 
photoemission from metals as previously shown for 
Cu(111)~\cite{krueger11}.

\begin{figure}[!htb]
\begin{center}
\includegraphics[clip=,width=0.99\columnwidth,height=6.0cm]{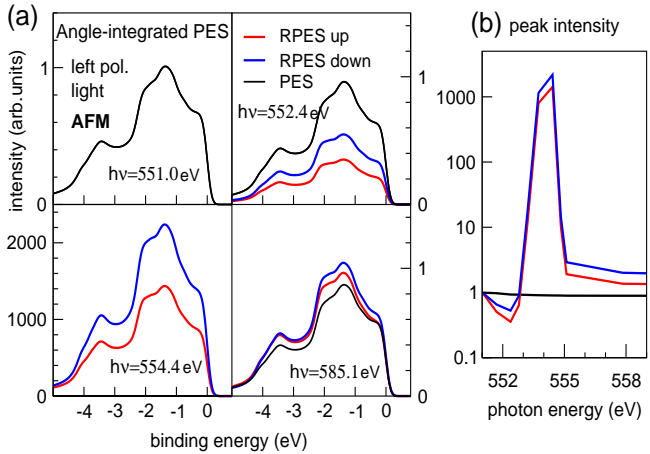}
\end{center}
\caption{a) Spin-resolved, angle integrated RPES and PES spectra
of AFM Cr(110) with circular polarized light incoming
 along the spin quantization axis [001] 
and photon energies across the $L_3$-edge resonance.
A gaussian broadening of 0.27~eV FWHM was applied.
Note the different intensity scale for $h\nu=554.4$~eV.
In PES, spin-up and down intensities are equal in all cases.
b) Maximum peak intensity as a function of photon energy.
}
\label{fig:conf2}
\end{figure}

Spin resolved, angle integrated PES and RPES spectra are shown in Fig.~\ref{fig:conf2} for the
AFM phase and several photon energies across the $L_3$-edge absorption
threshold. 
Left circular polarized light incident along the
magnetization axis [001] is considered. In this ``parallel'' geometry
the spectra, right polarized light produces the same spectra but with up and down spin exchanged. 
The maximum peak intensity as a function of photon energy is
plotted in Fig.~\ref{fig:conf2}b and shows the expected 
Fano profile.
The first photon energy (551.0~eV) is too low to excite the
core electron and so only direct PES is possible.
When the photon energy is raised to 552.4~eV, 
just below the absorption edge, direct and resonant process interfere
destructively, giving rise to the dip in the Fano profile.
Strong resonant enhancement is observed between 552 and 554.5~eV
(see e.g. the spectrum for 554.4~eV), which corresponds to transitions 
from the 2$p_{3/2}$ level into the unoccupied Cr 3$d$ band.
At $h\nu=585.1$~eV, well above threshold,
the resonant spectrum goes back to the non-resonant one.

The direct PES signal is non spin-polarized as expected for the AFM
phase. Appreciable spin-polarization is, however, found in RPES. 
This effect is here obtained for the first time
through first-principles calculations, and confirms the experimental finding in CuO~\cite{tjeng}, that in AFM systems RPES at the 2$p_{3/2}$-3$d$ resonance is spin-polarized when circular polarized light is used.

We now turn to angle and spin resolved spectra at maximum resonance 
(h$\nu$=554.4 eV), focusing on their four ``fundamental'' combinations (and their relation to local magnetic properties), constructed by different choices of photoelectron spin
($\uparrow$,$\downarrow$) and light helicity
($+,-$)$\equiv$(left,right): \\

\begin{tabular}{llll}
tot & $\equiv$ & $(\uparrow+) + (\uparrow-) + (\downarrow+) + (\downarrow-)$
 & (total) \\
spr & $\equiv$ & $(\uparrow+) + (\uparrow-) - (\downarrow+) - (\downarrow-)$
 & (spin-resolved) \\
dic & $\equiv$ & $(\uparrow+) - (\uparrow-) + (\downarrow+) - (\downarrow-)$
 & (dichroic) \\
mix & $\equiv$ & $(\uparrow+) - (\uparrow-) - (\downarrow+) + (\downarrow-)$
 & (mixed) \\
\end{tabular} 
\\

The ``mixed'' spectrum was the one considered in Refs~\cite{tjeng,sinkovic}
and claimed to be sensitive to local magnetic moments in non-ferromagnetic 
samples.

The normal emission RPES spectra (Fig~3a,b) (total spectra) for parallel geometry  
consist of a single 
peak at 0.8-0.9~eV binding energy, very similar to the low energy 
non resonant spectrum in Fig.1c ($\theta=0^o$).
AFM and PM spectra are almost identical except for a small shift 
of $\sim 0.1$~eV, which reflects the small exchange splitting 
of the AFM Cr-3$d$ bands.
The dichroic (dic) and spin-resolved (spr) signals vanish 
for both PM and AFM phase, as expected since the system is globally non-magnetic
in both cases, and the set up is non chiral.

However, the mixed signal is non-zero with a large amplitude
($\sim 1/3$ of total), in agreement with the experimental results in
AFM CuO~\cite{tjeng}. Surprisingly, we find a non-zero mixed signal
not only in the AFM, but also in the PM phase with nearly the
same intensity.
It is important to note that we are not considering a Curie paramagnet
(such as Ni above~$T_C$~\cite{sinkovic}) with disordered and/or
fluctuating magnetic moments, but a Pauli PM state, where the
magnetization is strictly zero in all points of space. Therefore, our
finding that the mixed signal is essentially unchanged when going
from the AFM to the PM state unambiguously proves that it is unrelated 
to local magnetic moments, in contrast to the interpretation in 
Refs~\cite{tjeng,sinkovic}.

\begin{figure}[!htb]
\begin{center}
\includegraphics[clip=,width=0.99\columnwidth]{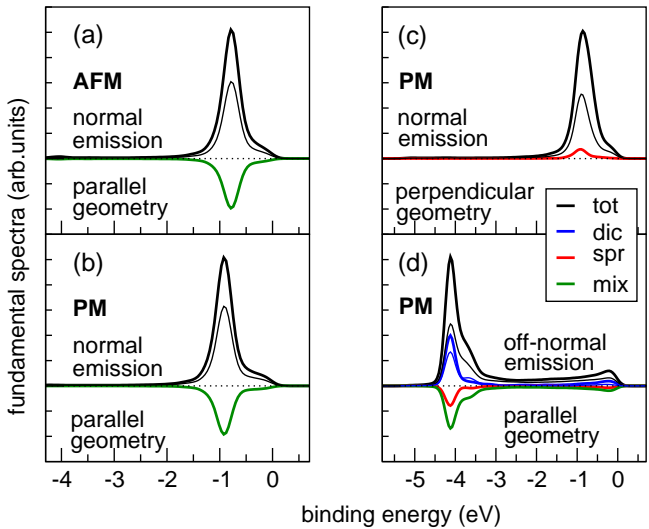}
\end{center}
\caption{
Angle-resolved fundamental spectra of Cr(110).
RPES as thick lines for $h\nu=544.4$~eV.
Normal (i.e., non resonant) ARPES as thin lines, intensity $\times 1000$.
All spectra are rescaled to equal peak height of RPES-tot.
(a) AFM, (b-d) PM phase.
(a-c) Normal emission. (d) Emission in the xy-plane, off-normal by 23$^o$
(vector {\bf e} in Fig.1~a).
Light incidence parallel (a,b,d) or perpendicular (c) to
spin-quantization axis~{\bf e}$_z$.
Light vector in (c) is shown as {\bf p} in Fig.1a.
}
\label{fig:conf3}

\end{figure}

Rather than being of magnetic origin, 
the non-zero mixed signal is in fact induced by angular momentum 
transfer from the light helicity to the electron spin via SO 
in the core shell together with a strong exchange effect in 
the decay process.
To see this, 
consider light with left (+) helicity and a non-magnetic ground state.
The 2$p_{3/2}$-3$d$ optical transition has a larger amplitude for
spin-up than for spin down electrons because of the dominantly
parallel alignment of spin and orbit in 2p$_{3/2}$. For example, for
an empty or spherically symmetric 3$d$ shell the intensity ratio is 5:3.
Consider now a spin-up electron transition. The RPES intermediate
state has one extra spin-up electron in the 3$d$-shell (denoted
$u$$\uparrow$) and a 2$p$-hole of dominant spin-up character.
This state decays through Coulomb interaction to the photoemission
final state with one 3$d$-hole and the photoelectron. The direct
Coulomb matrix elements is of the form
$\langle k\sigma,c$$\uparrow$$|V|v\sigma,u$$\uparrow$$\rangle$
which is independent of the photoelectron spin $\sigma$. So the
direct decay alone would lead to a spin-balanced photocurrent.  
For the exchange decay, the matrix element is 
$\langle k\sigma,c$$\uparrow$$|V|u$$\uparrow$$,v\sigma\rangle
\sim\delta(\sigma,$$\uparrow$$)$.  
This is roughly as a large as the direct Coulomb term for spin-up
electrons (the radial matrix elements are exactly the same) but it is
zero for spin-down electrons. Since the exchange matrix elements are
substracted from the direct terms in Eq.~(1), the transition
probability for spin-up electron emission is strongly reduced by the
exchange process. This shows that a core-valence transition of a
spin up electron leads, through autoionization,
to a strongly spin polarized photocurrent with a majority of
spin down electrons.
As mentioned before, left circular polarized light promotes
dominantly spin-up electrons in the 2$p_{3/2}$-3$d$ transition.
Therefore it produces a majority of spin-down photoelectrons.
Under the assumption of complete cancellation between direct Coulomb and 
exchange matrix elements for parallel spins and by neglecting the direct 
valence photoemission, the ratio of spin-down to spin-up photoelectrons 
is 5:3, which corresponds to a spin-polarization (ratio of mixed over 
total signal) of~$-1/4$. 
In angle integrated RPES at maximum resonance
(Fig.~2a, $h\nu$=554.4~eV) we find a SP of $-0.21$, 
in good agreement with such model estimation.
These values agree also well with the measured spin-polarization 
in CuO~\cite{tjeng} and Ni~\cite{sinkovic}, which is
10--40\% depending on binding energy. Our findings clarify the physical mechanism inducing the presence of the mixed signal in both phases, and point to a critical re-examination of experimental observations.

Interestingly, we find that, contrary to the previous set up, it is possible to have a net spin polarization signal on the PM phase. This is possible under appropriate geometrical conditions, and even with unpolarized light. Such SP can be of opposite sign and be due to different active mechanisms. In Fig.~3c, normal emission spectra are shown for light incident
along [1$\bar{1}$0], i.e.\ {\em perpendicular} to the
spin-quantization axis {\bf s}={\bf e$_z$} (perpendicular geometry).  
As before, the dichroic signal is zero, 
as light incidence ({\bf p}) and electron emission
vector ({\bf n}) lie in a mirror plane of the surface (see Fig.~1~a).
However, the set up (including spin resolution) is chiral,
since the three vectors {\bf p}, {\bf n} and {\bf s} form a
right-handed frame. Thus SO-induced SP cannot
be ruled out by symmetry and a small, positive SP (in this case transverse to the scattering plane) is
indeed observed in RPES, even for unpolarized light.  
A similar SP from PM surfaces for unpolarized light   
was theoretically predicted in direct PES \cite{tamura} 
in a relativistic approach and 
confirmed by experiments \cite{kirschner,heinz}.
It was ascribed to broken symmetry due to the off-normal light
incidence together with SO in the initial states and phase shift differences. We do not observe this effect in non-resonant PES
since the SO coupling in the Cr 3$d$ valence states is very weak and 
neglected here. However, for RPES, such SP has to be related to the dynamical 
SP studied in atomic physics, which is known to be related to phase shift differences in the final outgoing waves, and to be generally small \cite{spauger2,lohmann}. Our result confirms that such SP exists for an atom embedded in a solid and that it survives to the multiple scattering effects.

A SP signal in the PM phase is also present for parallel geometry with off-normal emission (Fig. 3d). In this case, the system composed by the surface, 
light incidence (along {\bf e$_z$}) and electron emission vector, 
is chiral. Therefore a dichroic signal is observed
even in non-resonant PES, known as circular dichroism 
in angular distribution~\cite{henk}.
In RPES, the angular momentum of the photon is partly transferred to
the electron spin through the SO coupling in the 2$p$ shell, leading to
non-zero intensity also for spin resolved and mixed signals. The spin
polarization is negative, i.e.\ photoelectrons are mainly polarized
antiparallel to their emission direction, because of the exchange
process in the autoionization decay. 
This finding suggests a Fano-like effect
in resonant processes for off
normal emission directions, which could be well studied 
along the same lines as direct PES on paramagnets \cite{prb63fano}.

In conclusion, we have presented a first-principles approach for
 RPES in solids and its application to Cr(110).
 By comparing Pauli PM and AFM states, we have shown that the
 mixed signal is essentially independent of local magnetic
 properties and we have clarified its origin: contrary to previous interpretations, this effect is induced by an angular momentum transfer from the photon to the electron spin, through SO coupling in
the core level and the exchange process in the autoionization decay. Our results show that caution must be taken in 
linking the spin polarized or mixed signal to local magnetic moments, 
all the more so as the photoelectron spin may have 
components along and across the light helicity. New effects in the SP suggest that a mapping of spin interactions in paramagnets and disordered magnetic structures could be obtained via full tomography experiments at the core resonances even with unpolarized light.

\end{document}